\documentclass[]{spie}  %>>> use for US letter paper
\usepackage{amsmath,amsfonts,amssymb}
\usepackage{graphicx}
\usepackage[colorlinks=true, allcolors=blue]{hyperref}
\usepackage{siunitx}
\usepackage[square,sort,comma,numbers]{natbib}
\usepackage{wrapfig}
\usepackage{aas_macros}

\title{\href{https://spie.org/AS24/conferencedetails/space-telescopes-and-instrumentation-optical-ir-mm-wave}{MoonLITE: a CLPS-delivered NASA Astrophysics Pioneers lunar optical interferometer for sensitive, milliarcsecond observing}}

\author[a]{Gerard T. van Belle}
\author[b]{David Ciardi}
\author[c]{Daniel Hillsberry}
\author[d]{Anders Jorgensen}
\author[e]{John Monnier}
\author[f]{Krista Lynne Smith}
\author[g]{Tabetha Boyajian}
\author[h]{Kenneth Carpenter}
\author[i,b]{Catherine Clark}
\author[j,k]{Gioia Rau}
\author[l]{Gail Schaefer}

\affil[a]{Lowell Observatory, 1400 West Mars Hill Rd., Flagstaff, AZ 86001, USA}
\affil[b]{NASA Exoplanet Science Institute, IPAC, California Institute of Technology, Pasadena, CA 91125, USA}
\affil[c]{Argo Space, Inc., 601 Cypress Ave, Suite 404, Hermosa Beach, CA, 90254, USA}
\affil[d]{New Mexico Institute Of Mining And Technology, 801 Leroy Pl., Socorro, NM 87801, USA}
\affil[e]{University Of Michigan, 500 S State St., Ann Arbor, MI 48109, USA}
\affil[f]{Mitchell Institute for Fundamental Physics \& Astronomy, Texas A\&M University, 576 University Dr., College Station, TX 77843, USA}
\affil[g]{Louisiana State University and A\&M College, Baton Rouge, LA 70803, USA}
\affil[h]{NASA Goddard Space Flight Center, 8800 Greenbelt Rd, Greenbelt, MD 20771, USA}
\affil[i]{Jet Propulsion Laboratory, California Institute of Technology, Pasadena, CA 91109, USA}
\affil[j]{National Science Foundation, 2415 Eisenhower Avenue, Alexandria, VA 22314, USA}
\affil[k]{NASA Goddard Space Flight Center, 8800 Greenbelt Road, Greenbelt, MD 20771, USA}
\affil[l]{The CHARA Array, Georgia State University, Mount Wilson, CA 91023, USA}

\authorinfo{Further author information: (Send correspondence to G.v.B.)\\G.v.B.: E-mail: gerard@lowell.edu, Telephone: +1 928 233 3207}

\begin{document} 
\maketitle

\begin{abstract} %200-300 words
MoonLITE (Lunar InTerferometry Explorer) is an Astrophysics Pioneers proposal to develop, build, fly, and operate the first separated-aperture optical interferometer in space, delivering sub-milliarcsecond science results.  MoonLITE will leverage the Pioneers opportunity for utilizing NASA’s Commercial Lunar Payload Services (CLPS) to deliver an optical interferometer to the lunar surface, enabling unprecedented discovery power by combining high spatial resolution from optical interferometry with deep sensitivity from the stability of the lunar surface.  Following landing, the CLPS-provided rover will deploy the pre-loaded MoonLITE outboard optical telescope 100 meters from the lander's inboard telescope, establishing a two-element interferometric observatory with a single deployment.  
MoonLITE will observe targets as faint as 17th magnitude in the visible, exceeding ground-based interferometric sensitivity by many magnitudes, and surpassing space-based optical systems resolution by a factor of 50$\times$.
The capabilities of MoonLITE open a unique discovery space that includes direct size measurements of the smallest, coolest stars and substellar brown dwarfs; searches for close-in stellar companions orbiting exoplanet-hosting stars that could confound our understanding and characterization of the frequency of Earth-like planets; direct size measurements of young stellar objects and characterization of the terrestrial planet-forming regions of these young stars; measurements of the inner regions and binary fraction of active galactic nuclei; and a probe of the very nature of spacetime foam itself.  
A portion of the observing time will also be made available to the broader community via a guest observer program.
MoonLITE takes advantage of the CLPS opportunity to place an interferometer in space on a stable platform -- the lunar surface -- and delivers an unprecedented combination of sensitivity and angular resolution at the remarkably affordable cost point of Pioneers.

%
% short abstract (1,200 characters, about above is 2,012)
%
%MoonLITE (Moon Lunar InTerferometry Explorer) is a an Astrophysics Pioneers proposal to develop, build, fly, and operate the first separated-aperture optical interferometer in space, delivering sub-milliarcsecond science results.  MoonLITE will leverage the Pioneers opportunity for utilizing NASA’s Commercial Lunar Payload Services (CLPS) to deliver an optical interferometer to the lunar surface, enabling unprecedented discovery power by combining high spatial resolution from optical interferometry with deep sensitivity from the stability of the lunar surface.  Following landing, the CLPS-provided rover will deploy the pre-loaded MoonLITE outboard optical telescope 100 meters from the lander's inboard telescope, establishing a two-element interferometric observatory with a single deployment.  MoonLITE will observe targets as faint as 17th magnitude in the visible, exceeding ground-based interferometric sensitivity by many magnitudes, and surpassing space-based optical systems resolution by a factor of 50$\times$. The capabilities of MoonLITE open a unique discovery space outlined briefly in the manuscript.

\end{abstract}


%\section{Summary of abstract (50-150 words)}

%\section{Speaker Bio (1000 characters)}

\section{Introduction}\label{sec:intro}  % \label{} allows reference to this section

Advances in spatial resolution have frequently led to unexpected discoveries in astrophysics.  For the first centuries of astronomy, those gains were largely limited to increasing aperture size of single-aperture telescopes.  In recent decades, however, the interferometric combination of telescopes has allowed astronomers to capture significant gains in spatial resolution at increasingly shorter wavelengths. Two recent, ground-breaking examples are the first direct observation of black holes by the Event Horizon Telescope \citep{EHT2019ApJ...875L...1E}, and the first detection of orbital precession of stars around the galactic center black hole by the VLTI-Gravity \citep{Gravity2020A&A...636L...5G}, which was part of the 2020 Nobel Prize in Physics. However, it is important to note that the VLTI observational capability in the optical involved multiple 8-meter-class telescopes at a facility costing hundreds of millions of euros.  Costly, large facilities are needed for terrestrial interferometry because the Earth's atmosphere limits integration times at those wavelengths to millisecond-scale durations. \textbf{Space-based optical interferometry does not have this sensitivity limitation, opening up a breadth of scientific avenues for relatively minimal complexity and modest cost.}   

\begin{wrapfigure}{r}{0.5\textwidth}
  \begin{center}
    \includegraphics[width=0.48\textwidth]{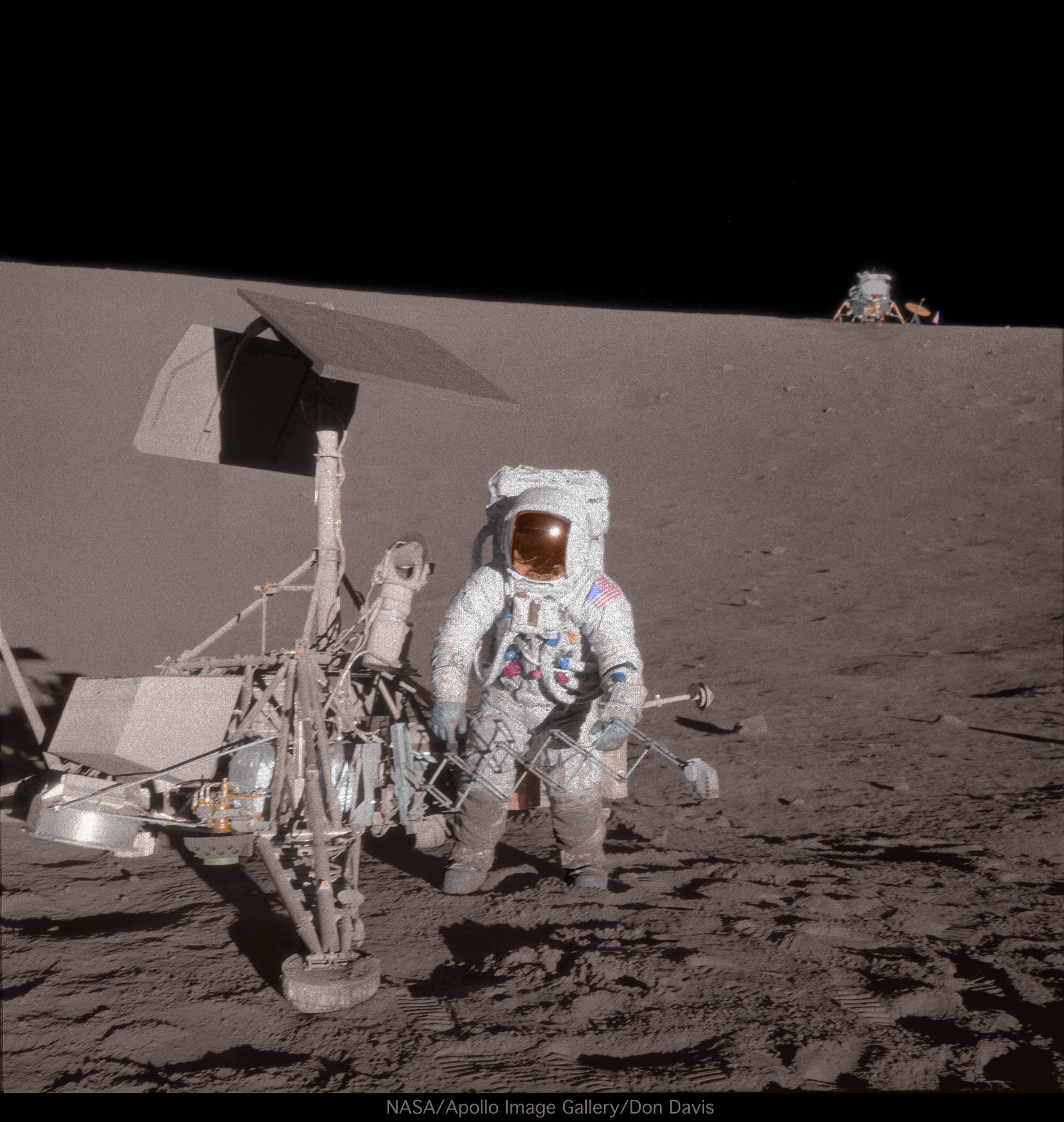}
  \end{center}
  \caption{Surveyor 3 being inspected by astronaut Charles Conrad; in the background is Apollo 12 Lunar Module \textit{Intrepid}.  Despite their derelict nature, these two spacecraft have remained precisely positioned relative to each other, and in absolute position on the lunar surface, for over 50 years.  (Credit: NASA/Don Davis).}\label{fig-S3A12}
\end{wrapfigure}

The Moon Lunar InTerferometry Explorer (MoonLITE) concept leverages NASA's newly flying Commercial Lunar Payload Services (CLPS) capability to deliver an optical interferometer to the lunar surface.  MoonLITE will enable unprecedented discovery power by combining the high spatial resolution of optical interferometry -- and deep sensitivity possible with the stability of the lunar surface -- at the remarkably affordable cost point of a NASA Astrophysics Pioneers.  Lunar surface siting yields the benefits of space (e.g., no atmosphere), but alleviates the need for complicated orbital spacecraft stability subsystems.

\section{Why the Moon?}

It has been suggested that, ``the only thing the moon has to offer astronomy is dirt and gravity'' \citep[][paraphrased]{Lester2007astro.ph..2437L}.  The discussion in \citep{Lester2007astro.ph..2437L} goes on to discuss the \textit{value proposition} that lunar siting has for astronomy.  It is useful to consider that value proposition when considering two important capabilities necessary for the optical interferometry technique.

First, the deceptively simple demands of \textit{telescope pointing} must be considered.  \citep{Lester2007astro.ph..2437L} argues that this is a solved problem because of examples like HST, which overlooks the considerable expense that has gone into this `simple' capability, and the considerable problems that even marquee efforts have had with telescope pointing.  The project landscape is littered with examples of missions that have terminated, failed, or degraded because of an inability to orient the spacecraft in a zero-gee environment -- starting with the US's first satellite, Explorer-1 \citep{Harland2005}, and more recently astrophysical facilities like HST, Kepler, Hitomi, WIRE, etc.  A 2009 Canadian Space Agency study of 156 failures worldwide between 1980 and 2005 listed the attitude control subsystem (ACS) as the leading cause of spacecraft failures \citep{Tafazoli2009}, causing almost a third of all failures, surpassing command \& data handling, power systems, telemetry, tracking \& command, structures \& mechanisms, and the payload. It is important to highlight that this is for all ACS failures, not just the even-more-demanding case of precision attitude control for telescopes.

Second, for an optical interferometer, the \textit{relative positions} of the multiple input telescopes are of equal importance.  To date, for orbital missions, there has been considerable effort -- and budget -- expended on developing formation flying capabilities; a particularly good summary is provided in \citep{Monnier2019BAAS...51g.153M}.  By and large, the current state of the art in formation flying -- which is still inadequate for optical interferometry -- means an investment of tens if not hundreds of millions of dollars.  
%Variations in the telescope positions has two impacts\footnote{At least two, that is.}: the change in pathlength on the corresponding interferometer arm must be compensated for at the nanometer level, and (somewhat more subtly) the effective sparse input aperture changes.

So why the moon?  The key here is that value proposition of lunar siting.  Let's consider the two spacecraft on the surface of the moon, Surveyor 3 and the Apollo 12 LM descent stage (Figure \ref{fig-S3A12}).  For the past 50 years, they have stayed 180 meters from each other, with their relative position remaining static at the sub-micron level, even though the two craft were never designed to maintain their relative positions.  Because of gravity, stationkeeping is free.

The Apollo 16 mission deployed the first lunar-based telescope, the Far Ultraviolet Camera/Spectrograph (UVC) \citep{Carruthers1972Sci...177..788C}.  Telescope pointing was achieved manually by the astronauts when starting an exposure; the astronauts then simply ignoring the telescope -- no gyros or star trackers needed.  \textbf{For both pointing and stationkeeping, gravity is not a bug, it's a feature}.

Finally, there is the question of dust.  Lunar dust is indeed nasty, even toxic, stuff.  Apollo astronauts reported congestion and fever, and similar dust on Earth has been associated with silicosis, asbestosis, and black lung disease.  The glass-shard nature of many of the dust particles, which have a peak size distribution at 0.3$\mu$m, leads to rapid frictional degradation of unprotected mechanical devices \citep{Park2006LPI....37.2193P}.   For astronauts returning to the lunar surface, mitigation of dust contamination will be of paramount importance.  However, for autonomous operations of properly designed and protected static optical systems, this will be less of a problem than anticipated, and mitigated through simple operational procedures.  This is not inferred from theory but from a TRL-9 existence proof: the Lunar-based Ultraviolet Telescope (LUT)  on board the Chang'e-3 lander \citep{Ip2014RAA....14.1511I} reported more than 1.5 years of highly stable operations on the lunar surface in 2015 \citep{Wang2015Ap&SS.360...10W}, and operated through 2018\footnote{``Thanks to power supplied by solar panels and a radioisotope heater unit, engineers think Chang’e-3 and LUT could have kept functioning for a few more decades, but at the end of 2018, China powered them down in order to focus on the fourth iteration of their Chang’e program.'' (\url{https://www.astronomy.com/observing/exploring-the-moon-chinas-change-missions/}, retrieved 2023-12-01.)}.  Simple operational procedures such as sealing off the LUT optics during lunar sunrise / sunset, when dust particles are expected to be levitated from the surface
\citep{Berg1976LNP....48..233B}, appear to have been effective in preventing degradation of the optics and the pointing mechanisms.  

Overall that value proposition of lunar siting is an advantageous context by which to examine the usefulness of astronomical observatories  on the lunar surface.  Our contention is that, particularly for interferometric telescope arrays, and within the rapidly evolving paradigm of increased surface access from NASA's Artemis program, the answer to that proposition has swung decisively in favor of the moon.

\section{Description of the Instrument}

\begin{figure}[h!]
%\begin{wrapfigure}{b,right}{11cm}
%\hspace{-1.75cm}
%\vspace{-0.5cm}
\includegraphics[width=1\textwidth, angle=0]{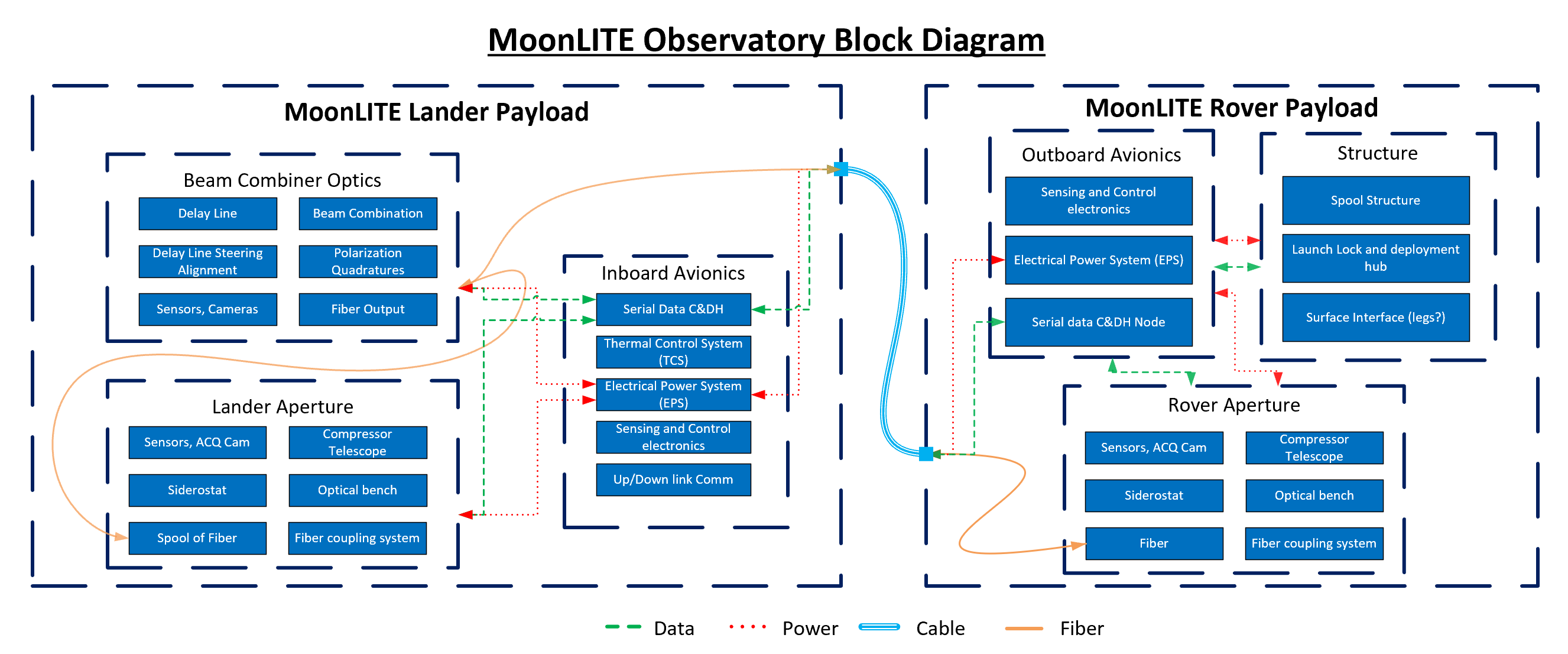}
\caption{\footnotesize MoonLITE subsystem block diagram.  Inboard and outboard apertures feed a backend beam combiner via fiber optics; the outboard aperture is connected back to the lander via an umbilical.}\label{fig-blockdiag}
%\end{wrapfigure}
\end{figure}

We developed the MoonLITE concept to rapidly take advantage of the now-flying CLPS lunar landers, and to demonstrate the viability and unique capability of optical interferometry from the surface of the moon.  We wanted to optimize the simplicity of MoonLITE, adopting known techniques from Earth-based arrays, and showing that even a pizza-box sized instrument can outperform industrial-level efforts back on Earth.  An overall block diagram of the instrument is seen in Figure \ref{fig-blockdiag}.

\textit{Telescope Feed Station Apertures.}
MoonLITE collects light on astronomical targets from two telescope feed stations.  One station, the `inboard', is mounted on the lander with the main instrument.  The second station, the `outboard', is deployed by the rover after landing, as described in the next section (\S \ref{sec_CONOPS}).  Both stations are identical.  A feed station has an articulating siderostat, which feeds a fixed 50~mm refractor telescope.  The telescope focuses light onto an image plane dichroic, which splits the $\lambda<620$~nm short-wavelength light off and feeds it to an image acquisition / fine guidance CCD.  The longer wavelength light passes through the dichroic and is coupled into a polarization-maintaining, single-mode fiber.

Siderostats have multiple advantages.  The articulating flat that feeds the telescope has a minimal swing radius, so the input aperture can be easily shuttered.  This is essential for preventing contamination of the optics due to dust levitation during lunar sunrise and sunset, as discussed previously.  Having a siderostat feed also means that the telescope, and any downstream optics such as the fiber feed, are fixed.  For the fiber in particular, a mechanically-fixed design means variations in throughput and beam rotation are eliminated.  Finally, a siderostat design means the system can be configured for retroreflection, and internal fringes can be measured.  This option allows for the measurement of internal, constant-term optical path difference (OPD).

\textit{Umbilical cable.} 
The umbilical line for the outboard telescope feed station includes an electrical connection for power and communications that is fed by the lander.  It also includes an optical fiber line for relaying observing light from the station back to the lander.  An additional optical fiber can also be considered for laser metrology-based pathlength measurement along the umbilical; this would allow for monitoring of temperature-induced fluctuations in OPD.  Such an option would need a metrology dichroic and retro-reflector added to the telescope feeds, as well as a fiber-based metrology source at the main instrument, both of which could be accommodated in straightforward manner.

The inboard telescope feed station has a simplified umbilical connection with the main instrument, consisting of an identical copy of the optical link.

In using an optical fiber link to relay the light from the outboard station back to the main instrument aboard the lander, two extremely important features are realized in the overall instrument architecture. First, scattered light is nearly eliminated, even during lunar daytime operations.  For architectures that involve free-space bulk optics relaying of light from outboard stations to a central combiner instrument, either lunar nighttime operations or significant amounts of baffling for the beam relay are required.  However, nighttime ops are likely a non-option in the near future given the extreme cold limiting battery life, the weight of sufficiently-capable batteries to combat cold and carry out ops, and the lack of immediately-available solar power.  

Secondly, with a fiber-based relay architecture, an operational two-element interferometer is obtained with the mechanical deployment of a single outboard station.  For free-space bulk optics relaying of light, to obtain a natural zenith-pointing of the instrument, two stations will need to be deployed at equal distances from a central combiner unit.  Such a deployment requirement doubles the number of mechanical actions involved in establishing the operational instrument, and potentially involves a rover CONOPS that requires returns to the lander to fetch a second outboard feed station.

\begin{wrapfigure}{r}{0.68\textwidth}
  \begin{center}
    \includegraphics[width=0.66\textwidth]{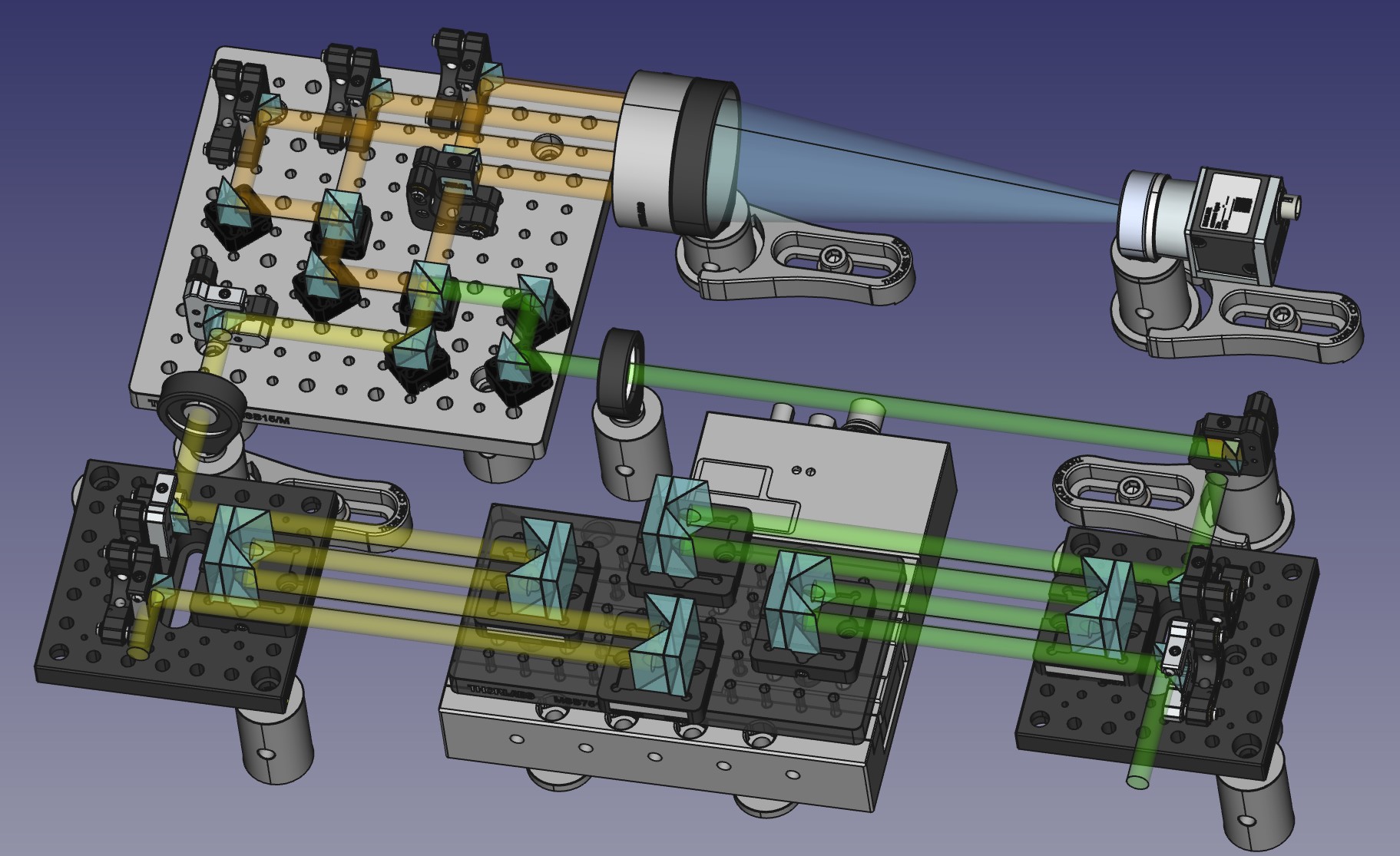}
  \end{center}
  \caption{A CAD rendering of the current lab test setup for the MoonLITE main instrument optical engineering development unit (opEDU).  Light from each of the two input beams (yellow and green) enters from the bottom, is routed through a multi-pass delay line, and is passed through a quarter waveplate (green beam) or compensator (yellow beam), meeting at a 50/50 beam splitter (where the beams becomes orange).  The combined beams are then split via polarizing beam splitters, creating four beams total, each of which encodes an individual fringe quadrature.  These four beams are focused onto a CCD camera.}\label{fig-lab_setup}
\end{wrapfigure}

\textit{Main Instrument: Delay Line.} 
Each of two input beams will be routed to the main instrument via fibers, will exit the fibers and be re-collimated, and fed into the instrument delay line.  The delay line has a multi-pass, dual-sided architecture such that the largest amount of optical path delay can be obtained with limited mechanical range.  For example, in our current lab test setup, each beam does a double back-and-forth pass to the delay line stage, and both beams follow such a path on opposite sides of the stage, meaning for a given mechanical stroke, the amount of optical stroke is 8$\times$ greater (eg. compare the yellow and green beams in Figure \ref{fig-lab_setup}).  The principal drawback of this architecture is that for a given amount of positioning jitter in the delay line linear actuator, the OPD jitter is similarly 8$\times$ greater.  Additionally, this increases the sensitivity to alignment errors of the beam path into the delay line stage.  Our current linear stage, a Zaber X-LDA075, has a 1~nm resolution linear encoder, and with 75~mm of travel, provides 600~mm of OPD in this configuration.  Additional passes in a flight instrument could increase the multiplicative effect even more, and every increase in stage travel adds to the available OPD in that multiplicative fashion.  In the case of MoonLITE, we intend to have $\sim$3 m of OPD delay for the 100~m baseline; the implications of this for sky coverage are presented in the CONOPS section (\S \ref{sec_CONOPS}).

\textit{Main Instrument: Combiner.} 
The beam combiner will utilize a polarization-based quadrature extraction architecture similar to the VLTI PRIMA instrument \citep{Sahlmann2008SPIE.7013E..1AS,Sahlmann2010SPIE.7734E..22S,Sahlmann2013A&A...551A..52S}, which has already been shown to perform in the visible in a flight-like packaging \citep{vanBelle2020SPIE11446E..2KV,vanBelle2022SPIE12183E..1DV}.  The two beams are combined after delay, and after passing through either a quarter-wave-plate or a compensator window, arrive at a 50/50 beam splitter.  The two output combined beams are then each routed to polarizing beam splitters.  The resulting four beams each photometrically encode one of the four fringe quadratures \citep[e.g., see Figure 6 of][]{Shao1988A&A...193..357S}. The four beams are focused onto a CCD detector.  The resulting measurement can be used to feedback control to the delay line.
Our current lab setup has not been optimized for size or weight, but its current size is already only 12''$\times$18'' (300mm$\times$450mm).

\begin{figure}
    \centering
    \includegraphics[width=0.95\textwidth]{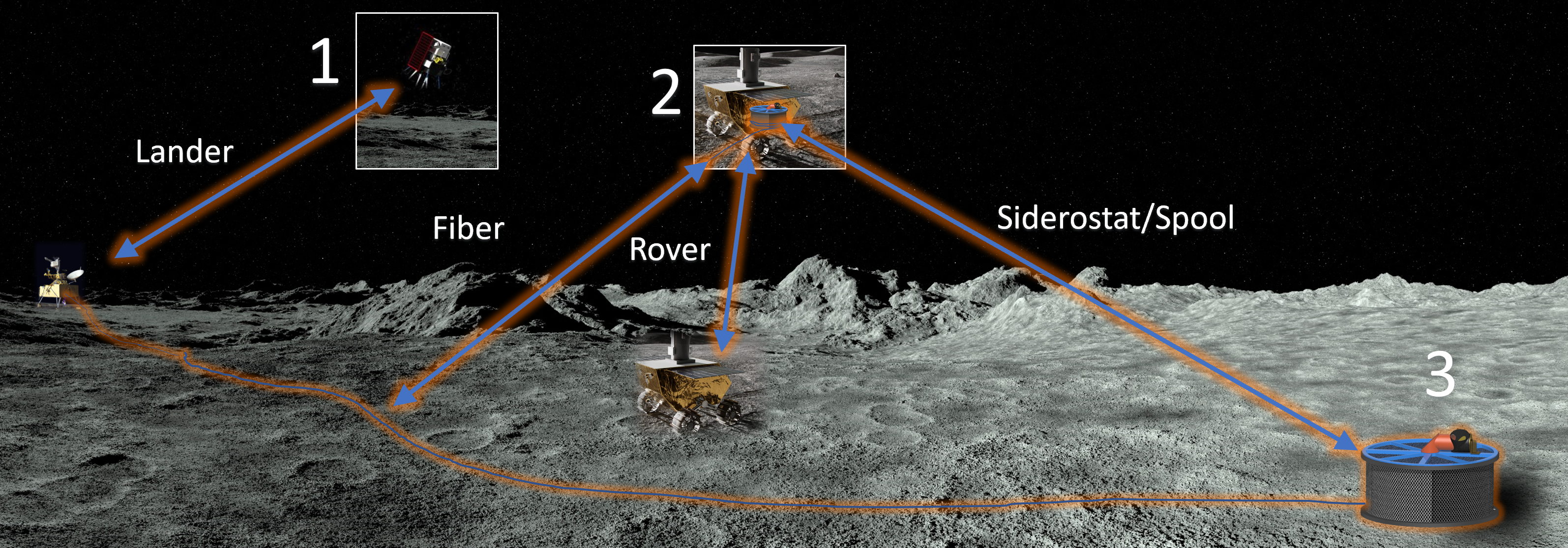}
    \caption{Concept of operations (CONOPS) for MoonLITE.  (1) A CLPS-provided lander arrives at the lunar surface.  (2) The CLPS-provided rover travels 100 meters away from the lander, unrolling a fiber umbilical.  (3)  The outboard siderostat station is deployed.  After calibration of the individual stations and the overall combined beams, science operations commence.}
    \label{fig:conops}
\end{figure}

\section{Concept of Operations (CONOPS)}\label{sec_CONOPS}

\textit{Physical deployment.} Prior to launch, the MoonLITE payload will be mounted into its payload bay aboard the CLPS lander.  This lander-mounted payload includes two principal components: (1) the main instrument, and (2) the inboard telescope feed station.  Additionally, the second outboard telescope feed station will be loaded onto the CLPS-provided rover prior to launch.  By pre-loading the outboard station onto the rover before launch, all failure modes associated with loading that station onto the rover after arrival at the lunar surface are eliminated.

The launch, cruise, and landing of the CLPS vehicle are all provided as a contracted service independent of MoonLITE.  Similarly, the deployment of the rover is also a CLPS-provided service.  Once the surface is reached, the first task of the rover will be to deploy the outboard station.  The rover will drive 100 meters away from the lander along an east-west line, unreeling (but not dragging) the umbilical cable back to the lander.  After reaching its desired distance from the lander, which has a flexible tolerance in distance and direction on the order of $\pm$1 meter, MoonLITE's single specific mechanical deployment happens: the rover sets down the outboard station and disengages from it.  The rover is now available to leave and conduct other CLPS tasks for other customers aboard the lander; the only restriction for the now-separate CONOPS of the rover is to avoid the umbilical cable.

The inboard station is optically identical to the outboard station, and also relays its light to the main instrument via a fiber. The inboard station fiber is coiled up inside the lander, meeting two important conditions. First, the light from each station has an identical amount of static optical path from its collector to the main instrument.  The means zero OPD -- eg. the natural pointing of the instrument -- will along the line running thru the zenith and perpendicular to the projection baseline vector of the two telescope stations onto the sky.  In the case of an east-west baseline, this natural pointing will run along the meridian.
Second, the light from each station encounters an identical amount of fiber -- as well as fiber-induced dispersion and other optical effects -- meaning the two beams will ultimately re-combine cleanly\footnote{Here we follow the second rule of interferometer beam relay.  The first rule is, ``Don't do anything bad to the light,'' and the second rule is, ``When you \textit{do} do something bad to the light, do the \textit{same} bad thing to all the beams.''}.

\textit{Calibration.}  Each telescope feed station will be activated individually, and will sweep out a series of images on the sky to determine the siderostat pointing models.  Astrometric recognition of the star images from the station acquisition camera will provide a mapping of the siderostat pointing to pointing coordinates for the sky.  Similar procedures and software already exist for Earth-based telescopes, have been tested with mobile interferometer stations at NPOI \citep{vanBelle2022SPIE12183E..04V}, and can recover a pointing model from a fully unknown state within 30 minutes.  

Once each individual station has a calibrated pointing model, an interferometric baseline vector will be determined.  MoonLITE will have each station point to the same bright astrometric reference object, and will commence looking for fringes.  Fringes seeking is carried out by examining the combined light from the two telescope stations and looking for the tell-tale sign of interference.  
Given the initial error envelope of the outboard station location relative to the inboard station, it is likely the initial search for fringes will be time-consuming, and may take many minutes.  However, once the first fringes are found, additional detections of other astrometric references will be much more rapid.  By expanding the detections into the full range of the available sky coverage of the instrument, a precise determination of the baseline vector of the instrument can be made.  This is the same procedure used at Earth-based facilities \citep{Lacour2014SPIE.9146E..2EL,Lacour2014A&A...567A..75L}, and typically results in a baseline determination at the tens of microns level.  As such, subsequent fringe detections only need to search this range of delay space, and are rapidly carried out.  

\begin{wrapfigure}{r}{0.5\textwidth}
  \begin{center}
    \includegraphics[width=0.48\textwidth]{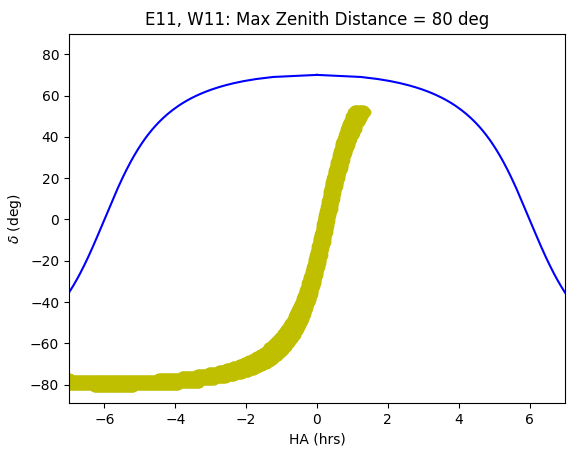}
  \end{center}
  \caption{Notional sky coverage chart for MoonLITE, given a putative landing latitude of $-20^o$, a 3-m delay line OPD range, and a 100 m baseline offset roughly $10^o$ from a east-west line. The offset results in a zone of extended sky coverage for targets at $\sim -77^o$, allowing for rotational aperture synthesis for those targets. Due to the rotational libration of the moon, the lunar axis `waggles' around by $6.7^o$; this effect is not accounted for in this plot. The net effect is that the edge of the coverage plot varies slightly with time, where depending on the libration phase, at the $\sim$10\% level regions outside the coverage will be accessible, and some inside the nominal coverage will not be accessible. The blue line represents the horizon.}\label{fig-sky_coverage}
\end{wrapfigure}

\textit{Operations.}  With the pointing models and baseline vector measured, science operations can commence.  By and large, as the moon rotates and objects cross the sky through the course of the lunar day of operations, they will become observable as they cross the range of sky coverage afforded by the delay line relative to the baseline vector (Figure \ref{fig-sky_coverage}).  For most objects on the sky, only a simple one-dimensional size measurement will be possible\footnote{The science cases for MoonLITE are tailored to objects for which this measurement is of extremely high value.}.  This is because, as noted in the instrument description, the nominal baseline layout for MoonLITE will be an east-west orientation, with only a narrow strip of sky coverage available.  Note that the units in Figure \ref{fig-sky_coverage} are hours of \textit{right ascension}; the amount of \textit{clock} time available for observing will be many hours, given the $\sim 28 \times$ slower rotation rate of the moon relative to the Earth.
Also, if a slight rotation of the baseline vector is selected with the deployment of the outboard station -- e.g., the rover heads slightly south of an exact east-west line -- a long tail in sky coverage can be dialed in a specific declination.  In Figure \ref{fig-sky_coverage}, a notional case is shown where many right ascension hours of coverage are possible at $\delta \sim 75^o$.  The intent here is that, if there are certain `marquee' targets available in a cluster at that declination range, a modicum of rotational aperture synthesis is now available to make two-dimensional maps of those targets\footnote{MoonLITE is rather `site agnostic', in that for any given landing latitude, the baseline can be tailored to maximize the available objects in its science case.}.  In this particular case, the Chamaeleon star formation region lies along this line of declination, and a science case for MoonLITE would be mapping the terrestrial regions of the young stellar objects.  

\textit{Survive-the-Night and the Following Day.}  A current requirement for CLPS is that the landers of that upcoming generation of surface hosts expect at least one cycle of non-operational `survive-the-night' and a second day of operations; payloads are expected to do the same, with a goal of up to six such cycles.  As such, our baseline required science plan for MoonLITE is two lunar days of science operations, with roughly 300 hours of science observing during each.  

\section{Science Cases}\label{sec_science}

The unique strength of MoonLITE will be its ability to observe objects at unprecedented sensitivity because of its substantially longer observing coherence times.  Unfettered by the Earth's atmosphere, our baseline requirement will be a 300 second exposure time, with a goal of 3000 seconds.  Based on this baseline coherence time, we estimate that MoonLITE will observe isolated objects down to a visual magnitude of 17.  This estimate is based on nominal numbers for fiber coupling efficiency \citep{Toyoshima2006JOSAA..23.2246T}, 2 e$^-$ detector read noise and 70\% quantum efficiency, $\sim$1-4 db/km fiber attenuation for commercially-available visible-light single-mode polarization-maintaining fibers, and 98.5\% mirrors reflectivity on 30 reflections (21 of which are used in a multiply-folded delay line).  This sensitivity outdoes current ground-based limits for direct fringe sensing by 6-7 magnitudes \citep{Lacour2019A&A...624A..99L}.  Techniques such as dual-beam feed can extend the sensitivity of Earth-based observatories, but these techniques are subject to the sky coverage necessity of a nearby fringe tracking reference object at the direct fringe sensing limit\footnote{The availability of IRS 16C with $m_K = 9.7$ within 1.2'' of Sgr A* for fringe tracking with VLTI-Gravity, and IRS 7 with $m_K = 6.5$ within 5.5'' of Sgr A* for AO phasing \citep{Gillessen2010SPIE.7734E..0YG} is a lightning-strikes-twice level of good fortune -- `Nature must want us to observe the galactic center.'}, as well as problematic noise and calibration problems \citep{gravity2022A&A...665A..75G}.  The 5 cm apertures of MoonLITE will outperform the sensitivity of 8 m Earth-based interferometer apertures.

As a two-element interferometer, the forte of MoonLITE science is ultra-high angular resolution observations in one principal axis.  This includes angular size measurements of astrophysical objects and detection of multiplicity along that axis.  In certain limited cases (e.g., \S \ref{sec-science_ysos}), a broader two-dimensional characterization of targets will be possible.  As MoonLITE is a high-precision, fiber-based instrument, we expect a visibility precision of 0.3\%  \citep{CoudeduForesto1997A&AS..121..379C,Scott2015PhDT.......699S}, which means angular sizes measurements down to 100$\mu$as should be possible with $<10\%$ errors, and measurements down to 0.4-1.4mas should be possible with $<0.1\%$ errors.

The following three examples are a small sample of the unexplored science vistas that are enabled by the unprecedented sensitivity of MoonLITE at sub-milliarcsecond scales in the visible; there are a substantial number of additional science investigations enabled by MoonLITE's unique capabilities.

\subsection{Radii of the Lowest-Mass Stars and Brown Dwarfs}\label{sec-science_mstars}

%\begin{figure}[h!]
\begin{wrapfigure}{t}{9.5cm}
%\hspace{-1.75cm}
\vspace{-0.25cm}
\includegraphics[width=0.54\textwidth, angle=0]{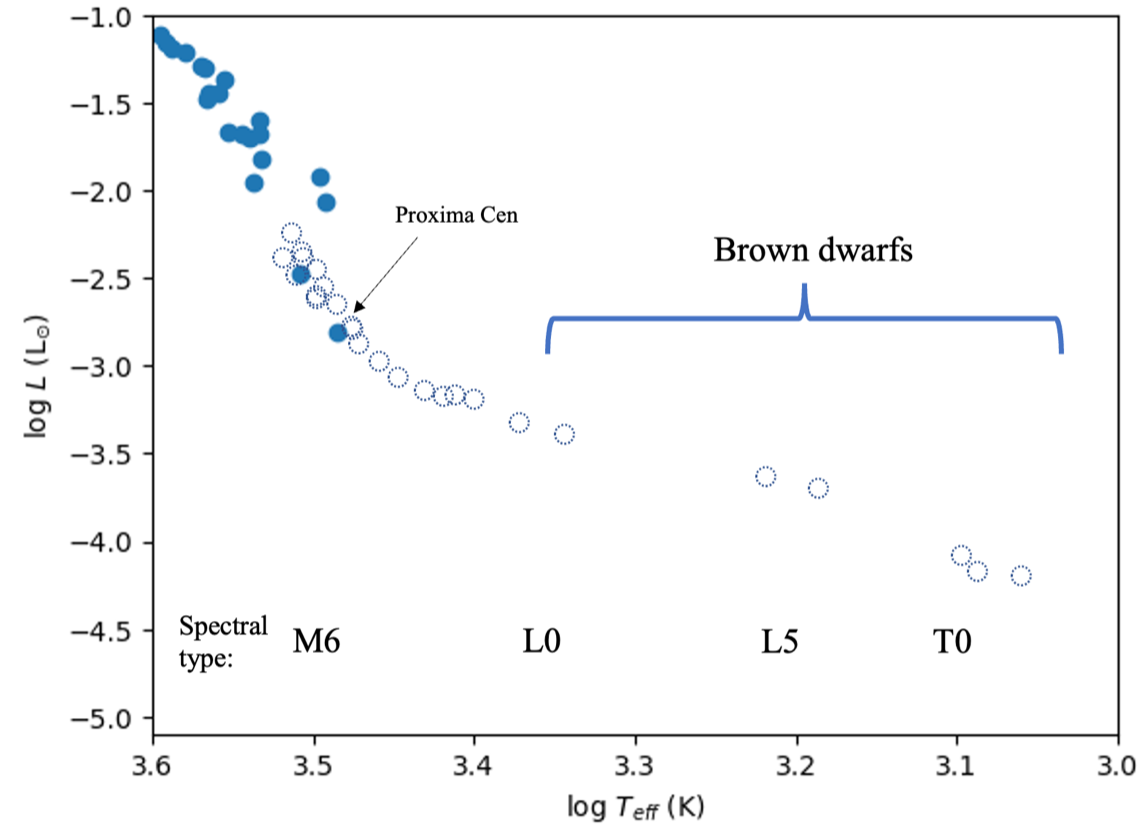}
\caption{\footnotesize 
Extending the empirical HR diagram for the coolest stellar and sub-stellar objects: MoonLITE will complete the census of M-dwarf stars from M5V to M9.5V, and will directly measure the sizes of multiple brown dwarfs.}\label{fig-M_BD_dwarfs}
\end{wrapfigure}
%\end{figure}

Direct measures of stellar angular sizes have been instrumental in \textit{empirically} establishing the reference linear radii and effective temperatures for main sequence \citep{Boyajian2012ApJ...746..101B,Boyajian2014AJ....147...47B} and giant stars \citep{vanbelle2021ApJ...922..163V}.  However, the sensitivity limits of ground-based measurements have restricted these essential calibrations of fundamental stellar parameters to main sequence stars earlier than the middle of the M-dwarf mass range \citep[eg. see Figure 2.1 in][]{vonBraun2017ephs.book.....V}. Empirical measurements are critical in this regime, where convection significantly complicates the models \citep{Berger2006ApJ...644..475B,vonbraun2014MNRAS.438.2413V}. As a result, there is a critical gap in our knowledge of the stellar radii and temperatures, and thus, the luminosities, for these stars and the planets that they host \citep{Schweitzer2019A&A...625A..68S}.  Furthermore, because of their faintness, interferometric measures of single M dwarfs truncate generally at M4V \citep{vonBraun2017ephs.book.....V}, with only a single measure of a M6V star (Prox Cen) in the literature \citep{Demory2009A&A...505..205D}. There are no direct measurements of later M dwarfs or sub-stellar brown dwarfs (Figure~\ref{fig-M_BD_dwarfs}). However, in the southern hemisphere, there are 13 M5-M6 stars, 11 M6.5-M9.5 stars, two L-dwarf systems (2MASSW J1507476-162738, a L5.5, and Luhman 16, a L7.5 + T0.5 pair), and a T1 brown dwarf \citep[$\epsilon$ Ind;][]{Reyle2021A&A...650A.201R} that are all both sufficiently bright ($m_I < 16.5$) and resolvable ($\theta > 0.11$ mas) by MoonLITE, with a comparable tally available in the north. By measuring the radii and habitable zone sizes of more than a dozen low-mass and ultra-cool dwarfs spanning the low-temperature spectral regime beyond the grasp of ground-based interferometers, MoonLITE can dramatically change our understanding of M stars and brown dwarfs.

\subsection{Young Stellar Objects}\label{sec-science_ysos}

\begin{wrapfigure}{right}{9cm}
%\hspace{-1.75cm}
%\vspace{-0.5cm}
\includegraphics[width=0.52\textwidth, angle=0]{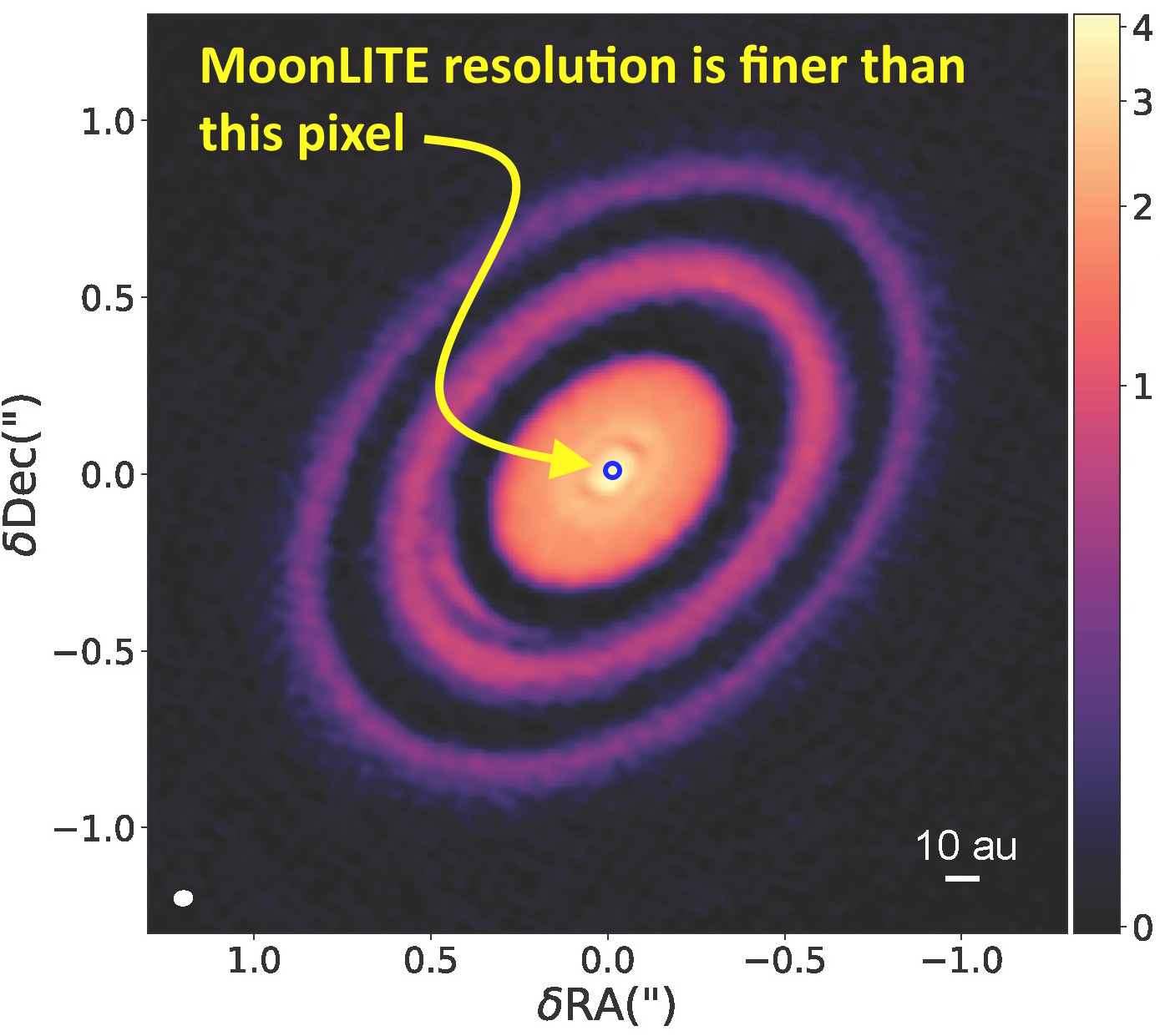}
\caption{\footnotesize The disk of HD163296 from ALMA \citep{Isella2018ApJ...869L..49I}, with an angular resolution of $40mas$. MoonLITE will have an angular resolution approximately 40$\times$ better, and will probe the terrestrial planet-forming region at 1 au.}
\label{fig:HD163296_disk}
\end{wrapfigure}

The discovery of thousands of exoplanets in a multitude of architectures has challenged our theories of planet formation.  Given the ubiquity of exoplanets, planet formation must be a highly-efficient process \citep{Burke2015ApJ...809....8B}, but theories that describe the formation and evolution of planets from protoplanetary disks around pre-main sequence stars have been poorly constrained because of a lack of sensitivity and resolution at the scales of planet formation. The 2020 Astrophysics Decadal Survey \citep{NASEM_Decadal_2021pdaa.book.....N} identified an understanding of the pathway to a habitable planet as one of the priority areas for the coming decade, and investigating how planets form and interact with their primordial disk and the pre-main sequence host star is a crucial element in understanding the morphology of our own Solar system and the diversity of exoplanetary systems that have been discovered \citep{Raymond2022ASSL..466....3R}.  These systems are beginning to be explored with sub-mm observations (Figure \ref{fig:HD163296_disk}), but MoonLITE can explore these YSOs in the visible, with higher resolution, for the first time.

%\begin{figure}[h!]
%\end{figure}

When young, more than half of the surface of a star may be covered in starspots \citep{paolino2023}, strongly altering energy transport in the outer layers, and causing stellar models to underestimate masses (and ages) by a factor of two \citep{somers2015}. Direct size measurements with MoonLITE will place significant constraints on the luminosity and mass, and thus models, of young stars. MoonLITE will also enable the first direct measurements of the inner regions of pre-main sequence stars, as well as sizes of the stars themselves.

The Chamaeleon I (Cha I) cloud contains no less than 80 young stellar objects (YSOs) with $m_V < 16$ and $\delta \sim -77$.  Given its near-pole declination and an optimized orientation of the baseline rotation, we expect significant baseline rotational sampling to be possible over the course of a lunar day (Figure \ref{fig-sky_coverage}). At a distance of 160 pc \citep{Feigelson2004ApJ...614..267F}, 1 mas corresponds to a scale of 0.16 au, providing ample resolution into the regions around the YSOs in this cloud.  Similarly, the star-formation region in Cepheus provides dozens of sufficiently bright objects at a high declination in the northern hemisphere \citep{Kun2008hsf1.book..136K}, although at slightly-less-preferable distances of 200 to 400 pc.

For simple size measurements of the central stars, a cursory review of known objects indicates many dozens of targets available all-sky with predicted sizes \citep{vanbelle1999PASP..111.1515V} greater than our 2.5\% error threshold % reference figure TBD in performance section?
at 0.2~mas.  Our planned YSO program will survey a dozen objects in detail for disk structure with $\geq$3 repeat observations, and more than two dozen YSOs for core object sizes. Inner disk structure and stellar binarity complicate the measurement of the sizes of the pre-main sequence stars, but observing enough position angles will enable simultaneous modeling of the diameters, inner disk structures, and binary separations for a subset of key sources.

\subsection{Active Galactic Nuclei}\label{sec-science_AGNs}
% {\color{blue} Writing assignment: Krista Lynne Smith }

\begin{figure}
%\hspace{-1.75cm}
%\vspace{-0.5cm}
\includegraphics[width=1\textwidth, angle=0]{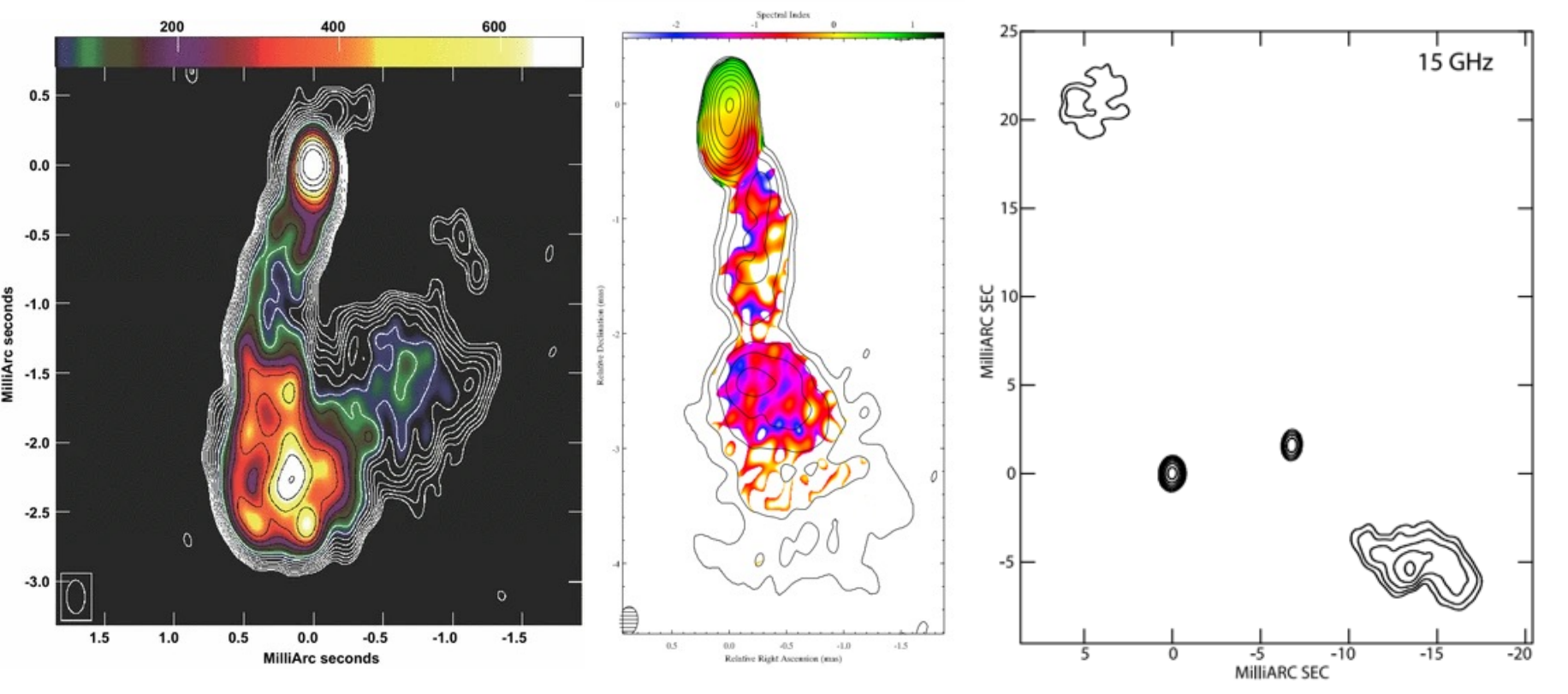}
\caption{\footnotesize Radio VLBI observations of the inner regions of the radio jets in 3C~84 and BL~Lacerta (left, center) and the binary AGN candidate in the radio galaxy 0402+379 (right). These figures are adapted from \cite{Nagai2014}, \cite{Gomez2016}, and \cite{Rodriguez2006}. The two jet structures span approximately 4~mas along the jet axis, which correspond to spatial scales of $\sim1.5$~pc and $\sim5$~pc for 3C~84 and BL~Lac, respectively; the binary separation is approximately 7~mas (7.3~pc).  MoonLITE will explore such objects at comparable scales in the optical for the first time.}
\label{fig:agn_examples}
\end{figure}

The ability of supermassive black holes to launch powerful relativistic jets that can far exceed the size of their host galaxies, and with profound effects on those galaxies' evolution, has been an astrophysical mystery for decades. The energy source of the jet may be the accretion disk or the spinning black hole itself, and the collimation mechanism may be strong, large-scale magnetic fields, interaction with the interstellar medium, or both. The activation mechanism of the accretion disk, and therefore the jet, occurs when material is funneled into the galactic center and onto the hole; however, the dominant mechanism for achieving this could be galaxy mergers, secular processes, or both. By far the dominant tool for studying the nuclear regions of the jet (often called the ``jet-base'') has been radio VLBI (Figure \ref{fig:agn_examples}), due to its ability to resolve on microarcsecond scales. Higher observing frequencies probe regions nearer to the central engine itself, and often reveal complex conical, cylindrical, or limb-brightened structures that improve our understanding of how the jet is constrained at very small spatial scales, and how it interacts with gas nearby.

A number of nearby radio galaxies exhibit parsec-scale radio structures that are misaligned with their larger, kiloparsec-scale radio jets. In some cases (e.g., NGC 1068), this radio emission is associated with a disk or torus that is roughly perpendicular to the larger structure. Studies have attempted to do this using proxy methods for studying the optical structure; for example, \cite{Lambert2021} found that the optical centroids of quasars from \emph{Gaia} coincide with downstream stationary radio features with high fractional polarizations in the jet, and that the optical emission on these scales arises from synchrotron emission in the jet.
%This same study finds a significant offset between the optical and X-band radio centroids in the opposite sense of that observed between the X-band and higher radio frequencies, indicating that the optical emission is dominated by downstream features like interactions with the ISM. 
%We propose, therefore, to use MoonLITE to 

%The question of the activation of merging binary supermassive black holes presents another pressing problem that is well studied in radio and mm-VLBI but missing a critical optical component. Theoretical investigations suggest that the type of accretion flows able to launch well-collimated radio jets (and thus appear as bright radio sources) are unlikely to generate significant UV-optical flux from a steady-state, optically thick but geometrically-thin accretion disk \citep{Narayan1994ApJ...428L..13N,Blandford2019}.

As supermassive black holes in merging galaxies approach from kiloparsec down to parsec-scale separations, when (or if) both black holes are activated by accretion and appear as a binary AGN is unclear. Until now, only radio interferometry was capable of measuring the binary fraction at the physical scales near 1 parsec, where the processes driving the merger are especially challenging theoretically (e.g., the ``final parsec problem'' \citep{Merritt2005LRR.....8....8M} of how the black holes lose sufficient angular momentum to enter the gravitational wave regime). Determining whether these same binaries appear in the optical as a double point source would be the first step in answering important questions about the accretion process onto the binary pair at various separations; is the accretion sufficient to power true, optically-bright accretion disks around the black holes, or only to power advection-dominated, low Eddington-ratio flows that are thought to power weak or young radio jets? 
%We therefore propose to observe a sample of 10-20 bright, parsec-scale candidate binaries known from radio VLBI observations. 
The MoonLITE AGN program will survey a dozen objects for detailed morphology with $\geq$3 repeat observations, and $\geq$a dozen AGN for core object sizes and binarity. MoonLITE will observe bright, very nearby radio galaxies with known parsec-scale jet structures that subtend a few to a few tens of milliarcseconds. These galaxies are well-studied by radio VLBI investigations at many frequencies and spanning a range of radio luminosities; candidates include 3C 84, Cygnus A, and NGC 4151 in the north, and PKS 0637-75, PKS 2153-69, NGC 6328, and Cen B in the south.  MoonLITE will directly study the structures of these galaxies in the optical -- on the same scales as in the radio -- for the first time. %We can change these sources around, or not mention specific sources, as needed. There will be bright radio galaxies in either hemisphere.

\section{Conclusion}

The MoonLITE instrument will demonstrate the viability and advantages of the lunar surface for astronomy, taking advantage of the rapidly-maturing surface access infrastructure of the CLPS landers available with in NASA Artemis program. MoonLITE will also present a unique opportunity to markedly extend the reach of high-resolution observing in the optical at unprecedented levels of sensitivity. Such improvements in the ability to image distant objects has always resulted in unexpected discoveries.

\acknowledgements

This research was carried out in part at the Jet Propulsion Laboratory, California Institute of Technology, under a contract with the National Aeronautics and Space Administration (80NM0018D0004).

\bibliographystyle{spiebib2b}
\bibliography{journal-references}

\end{document}